\documentclass[letterpaper, 10 pt, conference]{ieeeconf}
\IEEEoverridecommandlockouts
\usepackage{cite}

\usepackage{array,pdfsync,enumerate}
\usepackage[dvips]{epsfig}
\usepackage{amssymb,amsmath}
\usepackage{caption}
\usepackage{subcaption}
\usepackage{color}
\definecolor{linkcol}{rgb}{0,0.0,1.0}
\definecolor{citecol}{rgb}{0.0,0.6,0.0}

\usepackage[table,xcdraw]{xcolor}

\IEEEoverridecommandlockouts
\usepackage{cite}

\usepackage{amsmath,amssymb,amsfonts}
\usepackage{algorithmic}
\usepackage{graphicx}
\usepackage{textcomp}

\usepackage[pdfa]{hyperref}
\hypersetup{colorlinks,breaklinks,
	linkcolor=linkcol,urlcolor=citecol,
	anchorcolor=linkcol,citecolor=citecol,bookmarks=true}

\usepackage{caption}
\usepackage{subcaption}

\newtheorem{theorem}{Theorem}[section]

\newtheorem{remark}[theorem]{Remark}
\usepackage{graphicx}      
\usepackage{cite}

\begin{document}
	
	\title{Generation of Time-Varying Impedance Attacks Against Haptic Shared Control Steering Systems}
		\author{Alireza Mohammadi\textsuperscript{$\dagger$}, and Hafiz Malik
			\thanks{This work is supported by NSF Award 2035770. A. Mohammadi and H. Malik are with the
				Department of Electrical and Computer Engineering, University of Michigan-Dearborn, MI 48127 USA.   
				Emails: {\tt\small \{amohmmad,hafiz\}@umich.edu}. \textsuperscript{$\dagger$}Corresponding Author: A. Mohammadi.}%
		}
	
	\IEEEoverridecommandlockouts
	\makeatletter\def\@IEEEpubidpullup{6.5\baselineskip}\makeatother

\maketitle		
		
\begin{abstract}                
The safety-critical nature of vehicle steering is one of the main motivations for exploring the space of possible cyber-physical attacks against the steering systems of modern vehicles.  This paper investigates the adversarial capabilities for destabilizing the interaction dynamics between human drivers and vehicle haptic shared control (HSC) steering systems. In contrast to the conventional robotics literature, where the main objective is to render the human-automation interaction dynamics stable by ensuring passivity, this paper takes the exact opposite route. In particular, to investigate the damaging capabilities of a successful cyber-physical attack, this paper demonstrates that an attacker who targets the HSC steering system can destabilize the interaction dynamics between the human driver and the vehicle HSC steering system  through synthesis of time-varying impedance profiles. Specifically, it is shown that the adversary can utilize a properly designed non-passive and  time-varying adversarial impedance target dynamics, which are fed with a linear combination of the human driver and the steering column torques. Using these target dynamics, it is possible for the adversary to generate in real-time a reference angular command for the driver input device and the directional control steering assembly of the vehicle. Furthermore, it is shown that the adversary can make the steering wheel and the vehicle steering column angular positions to follow the reference command generated by the time-varying impedance target dynamics using proper adaptive control strategies. Numerical simulations demonstrate the effectiveness of such time-varying impedance attacks, which result in a non-passive and inherently unstable interaction between the driver and the HSC steering system.          
\end{abstract}
%
    \section{Introduction}
\label{sec:intro}

Haptic shared control (HSC), through which the driving task authority is balanced between the human driver and the driving assist system via an exchange of force at a motorized steering wheel, is becoming an indispensable part of advanced driver assistance systems (ADAS). In particular, HSC steering systems can achieve a common vehicle steering task objective between the automation and the human such as avoiding an obstacle or lane keeping. In these shared tasks,  the automation provides a continuous support to the human driver at the control level~\cite{marcano2020review}.  As demonstrated by Wang \emph{et al.}~\cite{wang2017effect}, haptic guidance can support drivers who have become passively fatigued during lengthy and monotonous trips.  Moreover, the perceived task load by the human driver can be reduced significantly during near-handling-limit driving scenarios as evidenced via the experiments by Katzourakis \emph{et al.}~\cite{katzourakis2014haptic}. In addition to guidance and reducing the driving perceived load, HSC steering systems can be used to make safe  transitions from humans to automation and vice versa~\cite{ludwig2018comparison,marcano2020review}. 

\begin{figure}[!t]
	\centering
	\includegraphics[width=0.35\textwidth]{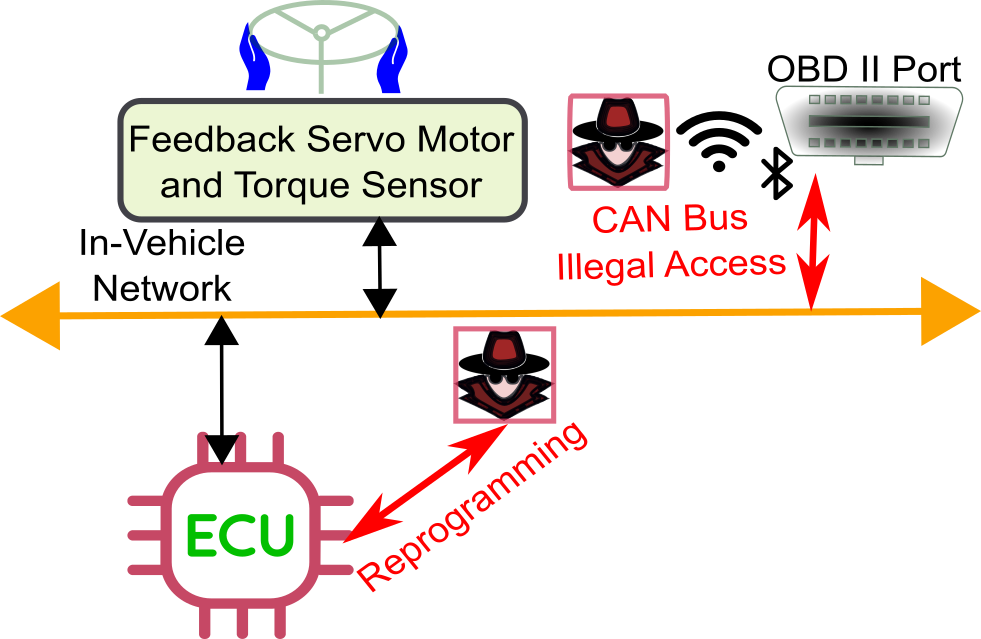} 
	\caption{\small{Several possible vectors through which cyber-physical attacks can be
executed against the 
			HSC steering system of a target vehicle.}}
	\vspace{-2ex}
	\label{fig:attackSteer}
\end{figure}

Given the safety-critical nature of steering, there is no wonder that cybersecurity researchers are interested in demonstrating cyber-physical vulnerabilities of vehicles through design of various attacks against their steering systems~\cite{kim2021cybersecurity} and analyzing the capabilities of an adversary who ``has made it to the last stage''~\cite{froschle2017analyzing} (see Figure~\ref{fig:attackSteer} for several possible cyber-physical attack vectors against the HSC steering system).

One of the main motivators for designing steering attacks can be traced back to the celebrated hacks by Miller and Valasek~\cite{miller2019lessons} in the field of vehicle cybersecurity. In a scentless manner and with leaving almost no forensic evidence behind, Miller and Valasek managed to steer a 2014 Jeep Cherokee into a ditch. Another hack against the active steering systems include the one proposed by Nekouei \emph{et al.}~\cite{nekouei2021randomized}, where the gains of the vehicle steering closed-loop controller are inferred by infiltrating the vehicular ad-hoc network. Chernikova \emph{et al.}~\cite{chernikova2019self}  have designed effective evasion attacks against  the deep neural networks that are used to predict the vehicle steering angle from images. Gao \emph{et al.}~\cite{gao2020distributed} have considered a class of CAN-based deception attacks against the steering mechanism of a four-in-wheel motor drive vehicle. 

\noindent{\textbf{Gap in the Literature.}} Despite the rich line of research on design of attacks against vehicle steering systems, no attack policy in the literature has directly focused on  destabilizing  interactions between the driver and the HSC steering system. Given the important role of human drivers in automated driving and the fact that up to \emph{Level 4 Automation}, humans should be capable of operating in driving conditions that are not supported by the automation (see, e.g.,~\cite{kyriakidis2019human}),  \emph{cyber-physical attacks that aim at destabilizing the human driver-vehicle interactions need to be thoroughly investigated}. 

\begin{figure}[!t]
	\centering
	\includegraphics[width=0.45\textwidth]{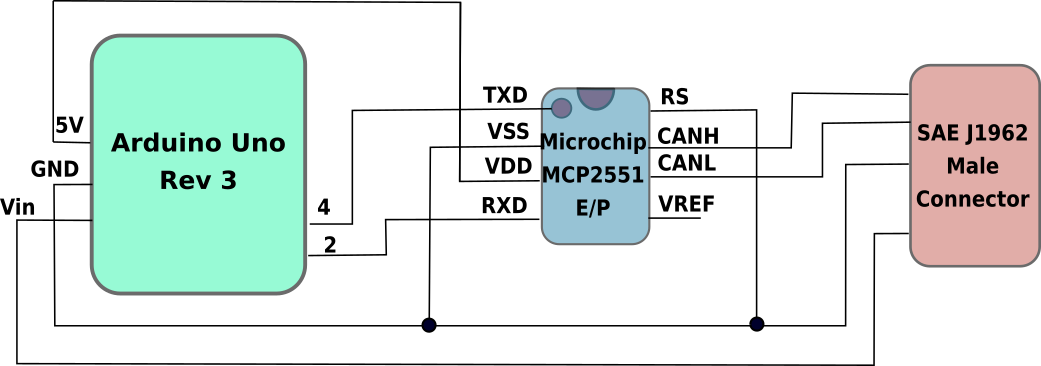} 
	\caption{\small{The inexpensive attacking device proposed by Palanca \emph{et al.}~\cite{palanca2017stealth} for accessing the CAN bus through the OBD-II port, which is central to the exchange of diagnostic messages within the in-vehicle network. Having access to the CAN bus, the adversary masquerades as a legitimate ECU and broadcast malicious messages to the target nodes~\cite{froschle2017analyzing}, e.g., the servomotors of the HSC steering system.}}
	\vspace{-2ex}
	\label{fig:palanaca}
\end{figure}

This paper investigates the adversarial capabilities for destabilizing the interaction dynamics between human drivers and HSC steering systems. We assume a cyber-physical attack scenario where the adversary has access to full disruption and full disclosure resources while having a limited knowledge of the HSC steering system parameters under attack (see, e.g., the inexpensive device depicted in Figure~\ref{fig:palanaca}  for accessing the controller area network (CAN) bus through the OBD-II port). We demonstrate that using a properly designed non-passive and  time-varying impedance target dynamics, which are fed with a linear combination of the human driver and the steering column torques, it is possible for the adversary to generate in real-time a reference angular command for the driver input device and the directional control steering assembly of the vehicle. Furthermore, we show that the adversary can make the steering wheel and the vehicle steering column angular positions to follow the reference command generated by the time-varying impedance target dynamics using proper adaptive control strategies while removing the need for the exact knowledge of the HSC steering system parameters.  

\noindent\textbf{Contributions of the Paper.} This paper contributes to the emerging field of cybersecurity in robotics and automation (see, e.g.,~\cite{khalil2021fault,Vilches22}) in several important ways. The first contribution of the paper is founded in using variable impedance control synthesis techniques (see, e.g.,~\cite{ferraguti2013tank,kronander2016stability}) for generation of cyber-physical closed-loop attacks against the HSC steering systems. In contrast to the traditional robotics literature, where the main objective is to render the human-robot interaction dynamics stable by ensuring passivity (see, e.g.,~\cite{buchli2011learning,ferraguti2013tank,kronander2016stability,ochoa2021impedance}), this paper takes the exact opposite route; namely, this paper proposes synthesizing time-varying impedance profiles that result in a non-passive interaction between the human driver and the vehicle HSC steering system. Second, to the best of our knowledge, no other attack policy has directly aimed at destabilizing  the interaction between the driver and the HSC steering system. Given the safety critical nature of HSC steering systems, which play an integral role in several applications such as seamless transition from automation to human and creation of a symbiotic driving experience, this paper contributes to understanding the physical damaging capabilities of adversaries who target HSC steering systems. 

%

The rest of this paper is organized as follows. First, we review the necessary preliminaries in Section~\ref{sec:dynModel}. Next, we present the time-varying impedance target dynamics and provide conditions under which these dynamics become non-passive in Section~\ref{sec:TargetDyn}. Thereafter,  in Section~\ref{sec:attackDesign}, we present  our closed-loop attack policy based on using an adaptive control scheme for making the steering wheel and the vehicle steering column angular positions to follow the reference command generated by the target dynamics. After validating our proposed attack policy through simulation results in Section~\ref{sec:sims}, we conclude the paper with further remarks, insights, and directions for future research in Section~\ref{sec:conc}. 
     \section{Preliminaries}

\subsection{Prior Literature on Impedance-Based Control of Haptic Shared Control Steering Systems}
Impedance control  is finding its way in the design of efficient HSC controllers in ADAS applications~\cite{chugh2018comparison}. The \emph{key idea in impedance control} for HSC steering systems is to  create a proper virtual inertia-spring-damper behavior resulting in a good steering feel and effective task sharing between the human driver and the automation. Izadi and Ghasemi~\cite{izadi2020adaptive,izadi2021modulation} have proposed an adaptive impedance control scheme to smoothly balance the authority of control in-between the human driver and his/her automation assistance system  through a motorized steering wheel. Chugh \emph{et al.}~\cite{chugh2020approach} have utilized open-loop driving maneuvers to design reference impedance models for the generation of appropriate haptic feedback for HSC steering systems in the linear range of vehicle handling dynamics. To ensure a stable interaction between the human driver and the steering wheel system, passivity techniques from the field of robotics teleoperation have also been utilized for designing controllers in HSC steering applications. For instance, Lee \emph{et al.}~\cite{lee2022haptic} have designed a haptic control algorithm with passivity guarantees for steering wheel systems to remove the possibility of having limit cycles and unstable behavior due to the unwanted effects of sampling and quantization.

\subsection{The Dynamics of the Steering System}
\label{sec:dynModel}
\begin{figure}[!t]
	\centering
	\includegraphics[width=0.35\textwidth]{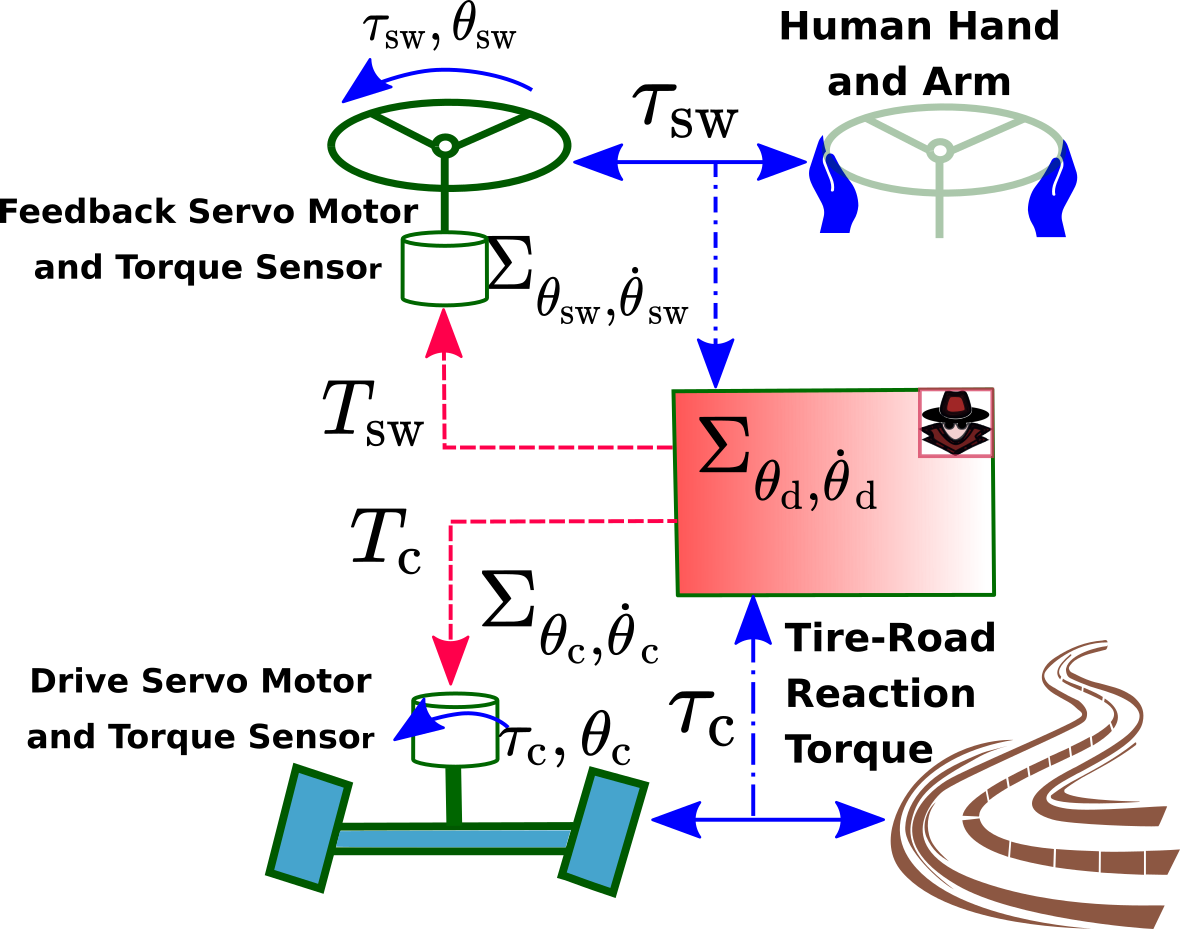} 
	\caption{\small{The HSC steering control system consisting of two dynamical control systems: (i) $\Sigma_{\theta_{\text{sw}}, 
				\dot{\theta}_{\text{sw}}}$, which governs the behavior of the driver input device (i.e., the steering wheel); and (ii) $\Sigma_{\theta_{\text{c}}, \dot{\theta}_{\text{c}}}$, which governs the behavior of the steering column. Using the proposed attack generation method in this paper, which relies on time-varying and non-passive target dynamics $\Sigma_{\theta_{\text{d}}, \dot{\theta}_{\text{d}}}$, the adversary generates two closed-loop attack inputs $T_\text{sw}$ and $T_\text{c}$, which destabilize the interaction dynamics between the driver and the HSC steering system.}}
	\label{fig:steerBlock}
\end{figure}

Sensors, ECUs, and servo motors form the backbone of HSC steering systems, which aim at transmitting the driver input to the servo motor mounted on the steering gear assembly in an electronic manner. In turn, for creating a proper `feel' of the road through haptic feedback to the human, the self-aligning reaction torques due to the tire--road interface forces are reflected back to the driver through the servo motor located on the steering wheel. The coordination between the torques transmitted to the steering column and the ones reflected back to the driver at the steering wheel are handled through the ECUs and proper control algorithms coded on them. Figure~\ref{fig:steerBlock} depicts a schematic block diagram of the HSC steering system investigated in this paper. 

The steering system dynamics, through which the driver/driving-assist-system interaction are governed,  can be written as (see, e.g.,~\cite{lazcano2021mpc,setlur2006trajectory}) 
\begin{subequations}
	\begin{align}
		\Sigma_{\theta_{\text{sw}}, 
			\dot{\theta}_{\text{sw}}}:\, I_{\text{sw}} \ddot{\theta}_{\text{sw}} + N_{\text{sw}}(\theta_{\text{sw}}, 
		\dot{\theta}_{\text{sw}}) &= \alpha_{\text{sw}} \tau_{\text{sw}} + T_{\text{sw}}, 
		\label{eq:dynSW}
		\\
		\Sigma_{\theta_{\text{c}}, 
			\dot{\theta}_{\text{c}}}:\,I_{\text{c}} \ddot{\theta}_{\text{c}} + N_{\text{c}}(\theta_{\text{c}}, 
		\dot{\theta}_{\text{c}}) &= \alpha_{\text{c}} \tau_{\text{c}} + T_{\text{c}}. 
		\label{eq:dynC}
	\end{align}
	\label{eq:dynMain}
\end{subequations}
\hspace{-2ex} In~\eqref{eq:dynMain}, the driver input device dynamics $\Sigma_{\theta_{\text{sw}},\dot{\theta}_{\text{sw}}}$ are given by~\eqref{eq:dynSW}, where  $\theta_{\text{sw}},\dot{\theta}_{\text{sw}},$ and $\ddot{\theta}_{\text{sw}}$, denote the driver input device angular position, its rate of change, and its associated angular acceleration, respectively. Furthermore, the inertia of the driver's input device is given by $I_{\text{sw}}$. The driver's arm is assumed to be rigidly coupled with the steering wheel and be exerting the torque $\tau_{\text{sw}}$ on it. Additionally, $T_{\text{sw}}$ is the control/attack input torque applied to the steering  wheel. Similarly, the steering column dynamics $\Sigma_{\theta_{\text{c}},\dot{\theta}_{\text{c}}}$ are given by~\eqref{eq:dynC}, in which $\theta_{\text{c}},\dot{\theta}_{\text{c}},$ and $\ddot{\theta}_{\text{c}}$, denote the steering column angular position, its rate of change, and its associated angular acceleration, respectively. The steering column dynamics $\Sigma_{\theta_{\text{c}},\dot{\theta}_{\text{c}}}$, whose inertia is given by $I_{\text{c}}$, is under the influence of the tire/road reaction torque $\tau_{\text{c}}$ and an additional control/attack input  $T_{\text{c}}$. The torque input scaling factors  $\alpha_{\text{sw}}$ and $\alpha_{\text{c}}$ are positive constants that might arise due to system gearing. Finally, the damping and friction effects are captured by the functions $N_{\text{sw}}(\cdot)$ and $N_{\text{c}}(\cdot)$, respectively. 

It is assumed that the damping and friction functions $N_{\text{sw}}(\cdot)$ and $N_{\text{c}}(\cdot)$ can be expressed as 
\begin{subequations}
	\begin{align}
	N_{\text{sw}}(\theta_{\text{sw}}, \dot{\theta}_{\text{sw}})  &= B_{{\text{sw}}} \dot{\theta}_{\text{sw}} + K_{{\text{sw}}} {\theta}_{\text{sw}},
	\label{eq:linFunc1}
	\\
	N_{\text{c}}(\theta_{\text{c}}, \dot{\theta}_{\text{c}})  &= B_{{\text{c}}} \dot{\theta}_{\text{c}} + K_{{\text{c}}} {\theta}_{\text{c}},
	\label{eq:linFunc2}
	\end{align}
	\label{eq:linFunc}
\end{subequations}
\hspace{-2ex} where $B_i$ and $K_i$, $i\in \{ \text{sw}, \text{c}\}$ are the damping and stiffness coefficients of the driver input device and the steering column, respectively. It can be seen that the functions in~\eqref{eq:linFunc1} and~\eqref{eq:linFunc2} can be written in the following linearly parametrizable forms 
\begin{subequations}
	\begin{align}
		N_{\text{sw}}(\theta_{\text{sw}}, \dot{\theta}_{\text{sw}})  &= Y_{N_{\text{sw}}}(\theta_{\text{sw}}, \dot{\theta}_{\text{sw}}) \phi_{N_{\text{sw}}},
		\label{eq:linParam1}
		\\
		N_{\text{c}}(\theta_{\text{c}}, \dot{\theta}_{\text{c}})  &= Y_{N_{\text{c}}}(\theta_{\text{c}}, \dot{\theta}_{\text{c}}) \phi_{N_{\text{c}}}. 
		\label{eq:linParam2}
	\end{align}
	\label{eq:linParam}
\end{subequations}
\hspace{-2ex} where $Y_{N_{i}}(\theta_{i}, \dot{\theta}_{i}):=[\theta_{i},\, \dot\theta_{i}]\in \mathbb{R}^{1\times 2}$, $i\in \{ \text{sw}, \text{c}\}$, is a regression matrix consisting of the measurable quantities $\theta_{i}$ and  $\dot{\theta}_{i}$. Furthermore, $\phi_{N_{i}}:=[K_{i},\, B_{i}]^\top\in \mathbb{R}^{2\times 1}$, $i\in \{ \text{sw}, \text{c}\}$, is a constant vector of the damping and stiffness coefficients. 

As it will be shown in Section~\ref{sec:attackDesign}, the linearly parametrizable form in~\eqref{eq:linFunc} will be used for designing adaptive attack policies and updating the unknown parameters of the driver/driving-assist-system dynamics  according to properly designed adaptive update laws.    
     \section{Adversarial Time-Varying Impedance Target Dynamics}
\label{sec:TargetDyn}
In this section we present a non-passive and  time-varying impedance target dynamics, which are fed with a linear combination of the human driver and the steering column torques from the torque sensor readings of the HSC steering system. We will demonstrate that it is possible for the adversary to generate in real-time reference angular commands for the driver input device and the directional control steering assembly of the vehicle where the target dynamics (i.e., the system $\Sigma_{\theta_{\text{d}}, \dot{\theta}_{\text{d}}}$ in Figure~\ref{fig:steerBlock}) become a non-passive system. 

Consider the following time-varying target dynamics 
\begin{equation}
\Sigma_{\theta_{\text{d}}, \dot{\theta}_{\text{d}}}: I_T \ddot{\theta}_{\text{d}} + N_T(\theta_{\text{d}}, \dot{\theta}_{\text{d}},t) = \alpha_{T_{\text{sw}}} \tau_{\text{sw}} + \alpha_{T_{\text{c}}} \tau_{\text{c}}, 
\label{eq:refModel}
\end{equation}
where $\theta_{\text{d}}$, $\dot{\theta}_{\text{d}}$, and $\ddot{\theta}_{\text{d}}$ are the target angular position, its rate of change, and the angular acceleration, respectively. The time-varying function $N_T(\theta_{\text{d}}, \dot{\theta}_{\text{d}},t)$ represents an auxiliary target dynamic function, consisting of time-varying damping and stiffness terms; and, the design parameters $\alpha_{T_{\text{sw}}}$ and $\alpha_{T_{\text{c}}}$ are scaling constants that weigh the driver's input torque $\tau_{{\text{sw}}}$ and the reaction torque $\tau_{{\text{c}}}$ from the steering column. 

\noindent{\textbf{The Adversarial Design Objective.}} The adversarial design objective for the auxiliary function $N_T(\theta_{\text{d}}, \dot{\theta}_{\text{d}},t)$ is to inject energy into the dynamical system $\Sigma_{\theta_{\text{d}}, \dot{\theta}_{\text{d}}}$ to create an unstable behavior for the reference trajectories $\theta_{\text{d}}$, $\dot{\theta}_{\text{d}}$, and $\ddot{\theta}_{\text{d}}$, which get modulated in response to the driver's input torque $\tau_{{\text{sw}}}$ and the reaction torque $\tau_{{\text{c}}}$ from the steering column. As demonstrated later in Section~\ref{sec:attackDesign}, these trajectories will be fed to an adaptive attack policy that will make the angular positions $\theta_{\text{sw}}$ and $\theta_{\text{c}}$ to follow $\theta_{\text{d}}(t)$ closely. 

\begin{figure}[!t]
	\centering
	\includegraphics[width=0.28\textwidth]{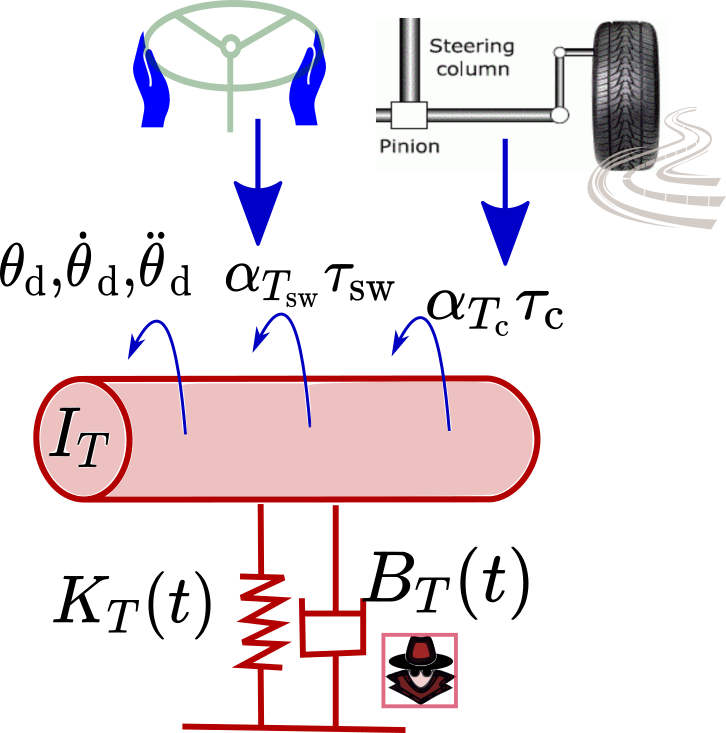} 
	\caption{\small{The target dynamics designed by the adversary utilize the torque inputs $\tau_{\text{sw}}$ and $\tau_{\text{c}}$ to generate malicious reference commands $\theta_{\text{d}}$, $\dot{\theta}_{\text{d}}$, and $\ddot{\theta}_{\text{d}}$. The torques  $\tau_{\text{sw}}$ and $\tau_{\text{c}}$ are fed into an impedance dynamics with variable stiffness $K_T(t)$ and variable damping $B_T(t)$, which are designed to be non-passive and aiming at energy injection into the driver/HSC-steering-system interaction dynamics via the malicious reference commands $\theta_{\text{d}}(t)$, $\dot{\theta}_{\text{d}}(t)$, and $\ddot{\theta}_{\text{d}}(t)$.}}
	\vspace{-2ex}
	\label{fig:steerTargetDyn}
\end{figure}

To design the adversarial target dynamic function $N_T(\theta_{\text{d}}, \dot{\theta}_{\text{d}},t)$, we consider 
\begin{equation}
N_T(\theta_{\text{d}}, \dot{\theta}_{\text{d}},t) = B_T(t) \dot{\theta}_{\text{d}} + K_T(t) {\theta}_{\text{d}}, 
\label{eq:NTtarget}
\end{equation}
where  $B_T(\cdot)$ and $K_T(\cdot)$ are two continuously differentiable time-varying stiffness and damping profiles. Therefore, the \emph{adversarial target dynamical system} takes the form (see, also, the block diagram in Figure~\ref{fig:steerTargetDyn})
\begin{equation}
I_T \ddot{\theta}_{\text{d}} + B_T(t) \dot{\theta}_{\text{d}} + K_T(t) {\theta}_{\text{d}} = \alpha_{T_{\text{sw}}} \tau_{\text{sw}} + \alpha_{T_{\text{c}}} \tau_{\text{c}}.
\label{eq:targetMain}
\end{equation}
\begin{remark}
	In contrast to the conventional robotics literature (see, e.g.,~\cite{buchli2011learning,ferraguti2013tank,kronander2016stability,ochoa2021impedance}), where the goal is to improve the robotic task performance by designing proper variable stiffness and damping profiles while ensuring stability, the goal of the adversary in this paper, who is targeting the vehicle HSC steering system, is to create an unstable behavior by inducing non-passive dynamics through a proper choice of the variable damping  $B_T(\cdot)$ and impedance $K_T(\cdot)$ functions.\hfill$\blacklozenge$ 
	\label{rem:timeVarying}
\end{remark}

\noindent{\textbf{The Adversarial Target Dynamics Storage Function.}} Let us consider the storage function, which was originally proposed by Kronander and Billard~\cite{kronander2016stability} for studying stability of time-varying impedance controllers (see, also, Remark~\ref{rem:timeVarying}), 
\begin{equation}
V_a(\theta_{\text{d}}, \dot{\theta}_{\text{d}},t) = \frac{1}{2} I_T (\dot{\theta}_{\text{d}}+ \alpha \theta_{\text{d}})^2 + \frac{1}{2} \beta(t) \theta_{\text{d}}^2, 
\label{eq:lyapCandid}
\end{equation}
for the target dynamics in~\eqref{eq:targetMain}. The  positive constant $\alpha$ in the storage function given by~\eqref{eq:lyapCandid} needs to be chosen in a way that the inequality  
\begin{equation}
\beta(t) := K_T(t) + \alpha B_T(t)  - \alpha^2 I_T \geq 0, 
\label{eq:betaFunc}
\end{equation}
holds for all $t>0$. 

\begin{remark}
	In addition to the storage function $V_a(\cdot)$ given by~\eqref{eq:lyapCandid}, there are other possibilities for defining candidate storage functions. For instance, consider the normalized unforced adversarial target dynamical system $\ddot{\theta}_d + \tfrac{B_T(t) }{I_T} \dot{\theta}_{\text{d}} + \tfrac{K_T(t)}{I_T} {\theta}_{\text{d}}=0$. It is possible to define a \emph{time-dependent Hamiltonian} as another potential candidate storage function for the normalized target dynamics. Defining the canonical momentum to be $p:=\zeta(t)^2 \dot{\theta}_d$, where $\zeta(t) := C_1 \exp(\tfrac{\int B_T(t)dt}{2I_T})$, the time-dependent Hamiltonian of these unforced dynamics are given by (see, e.g.,~\cite{kanasugi1995systematic} for further details) $H(\theta_d, p, t) = \frac{1}{2} \Big\{ \frac{p^2}{\zeta^2(t)} + \zeta^2(t) \frac{K_T(t)}{I_T} \theta_d^2 \Big\}$.\hfill$\blacklozenge$ 
	%
	%
	\label{rem:timeDep}
\end{remark}

Taking the derivative of the storage function in~\eqref{eq:lyapCandid} along the trajectories of the target dynamics~\eqref{eq:targetMain}, we arrive at 
\begin{equation}
\begin{aligned}
\dot{V}_a &= \big\{ \alpha I_T - B_T(t) \big\} \dot{\theta}_{\text{d}}^2 + \big\{ \frac{1}{2} \dot{K}_T(t) + 
 \frac{\alpha}{2} \dot{B}_T(t) \\ 
 & - \alpha K_T(t) \big\} \theta_{\text{d}}^2 +  \dot{\theta}_{\text{d}} \big( \alpha_{T_{\text{sw}}} \tau_{\text{sw}} + \alpha_{T_{\text{c}}} \tau_{\text{c}} \big). 
\end{aligned}
\label{eq:lyapCandidDeriv}
\end{equation}

It can be easily seen that if the time-varying stiffness $K_T(t)$, the time-varying damping $B_T(t)$, and their rates of change $\dot{K}_T(t)$ and  $\dot{B}_T(t)$ satisfy the inequalities 
\begin{subequations}
	\begin{align}
	\alpha I_T - B_T(t)  > 0,
	\label{eq:ineq1}
	\\
	\frac{1}{2} \dot{K}_T(t) + \frac{\alpha}{2} \dot{B}_T(t) - \alpha K_T(t) > 0,  
	\label{eq:ineq2}
	\end{align}
	\label{eq:targetIneq}
\end{subequations}
\hspace{-2ex} then the rate of change of the storage function $V_a(\cdot)$ satisfies the inequality 
\begin{equation}
\dot{V}_a > \dot{\theta}_{\text{d}} \big( \alpha_{T_{\text{sw}}} \tau_{\text{sw}} + \alpha_{T_{\text{c}}} \tau_{\text{c}} \big) + \frac{\rho_1}{2} {\theta}_{\text{d}}^2+ \frac{\rho_2}{2} \dot{\theta}_{\text{d}}^2,     
\label{eq:nonPassiveProof}
\end{equation}
where $\rho_1 = \inf\limits_{t>0} \big( \alpha I_T - B_T(t) \big)$ and $\rho_2 =  \inf\limits_{t>0} \big( \tfrac{1}{2} \dot{K}_T(t) + \tfrac{\alpha}{2} \dot{B}_T(t) - \alpha K_T(t) \big)$, for all $(\theta_{\text{d}}, \dot\theta_{\text{d}})\neq (0,0)$ and $t>0$.  which  renders the target dynamics in~\eqref{eq:targetMain} \emph{non-passive} with respect to the input-output pair $(\dot{\theta}_{\text{d}},\, \alpha_{T_{\text{sw}}} \tau_{\text{sw}} + \alpha_{T_{\text{c}}} \tau_{\text{c}})$. The adversary can utilize the weighting factors  $\alpha_{T_{\text{sw}}}$ and $\alpha_{T_{\text{c}}}$ as design variables. For instance, if the adversary is interested in creating a non-passive target dynamics in response to the driver input, they can choose  the driver input weighting factor $\alpha_{T_{\text{sw}}}$ to be positive and the steering column weighting factor $\alpha_{T_{\text{c}}}$ to be zero.

We remark that the inequalities given in~\eqref{eq:ineq1} and~\eqref{eq:ineq2} arise due to the particular choice of the storage function in~\eqref{eq:lyapCandid}, where mixed angular velocity and position terms appear in the derivative of the storage function along the reference dynamic system trajectories in~\eqref{eq:targetMain}. As it can be seen from~\eqref{eq:lyapCandidDeriv}, the choice of the storage function in~\eqref{eq:lyapCandid} results in the inequalities given by~\eqref{eq:ineq1} and~\eqref{eq:ineq2} that relate the time-varying damping, the time-varying stiffness, and their rates of change to ensure a non-passive behavior from the target dynamical system in response to the driver's input $\tau_{\text{sw}}$ and the steering column reaction torque $\tau_{\text{c}}$.

\noindent{\textbf{Design of the Adversarial Target Dynamics Impedance Profiles.}} To find conditions on the time-varying stiffness $K_T(t)$ and damping $B_T(t)$ profiles, which satisfy the inequalities given by~\eqref{eq:betaFunc},~\eqref{eq:ineq1}, and~\eqref{eq:ineq2}, we define 
\begin{equation}
	\kappa(t) := -K_T(t) - \alpha B_T(t) + \alpha^2 I_T. 
	\label{eq:kappa}
\end{equation}

Using the definition given by~\eqref{eq:kappa}, it can be seen that $\kappa(t)\leq 0$ due to~\eqref{eq:betaFunc} and  $-\tfrac{1}{2\alpha}\dot\kappa - K_T(t)> 0$ due to~\eqref{eq:ineq2}. Adding the sides of the inequality in~\eqref{eq:ineq1} to $-\tfrac{1}{2\alpha}\dot\kappa - K_T(t)> 0$ yields
\begin{equation}
	\dot{\kappa} <  2\alpha \kappa(t).  
	\label{eq:kappaIneq}
\end{equation}

From a straightforward application of the Gr\"{o}nwall-Bellman inequality (see, e.g.,~\cite[Lemma A.1, p. 651]{khalil2002nonlinear}) to~\eqref{eq:kappaIneq}, it can be seen that $\kappa(t) < \kappa(0) \exp(2\alpha t)$ holds for all $t>0$. Therefore, the definition in~\eqref{eq:kappa} yields the following constraint on the time-varying stiffness and damping profiles 
\begin{equation}
	K_T(t) +  \alpha B_T(t) >  \alpha^2 I_T + \big( K_{0} + \alpha B_{0} - \alpha^2 I_T \big) \exp(2\alpha t), 
	\label{eq:finalConst}
\end{equation}
where $B_{0} = B_T(0)$ and $K_0 = K_T(0)$ for all $t>0$.

\begin{figure}[!t]
	\hspace{-4ex}
	\includegraphics[height=0.27\textwidth]{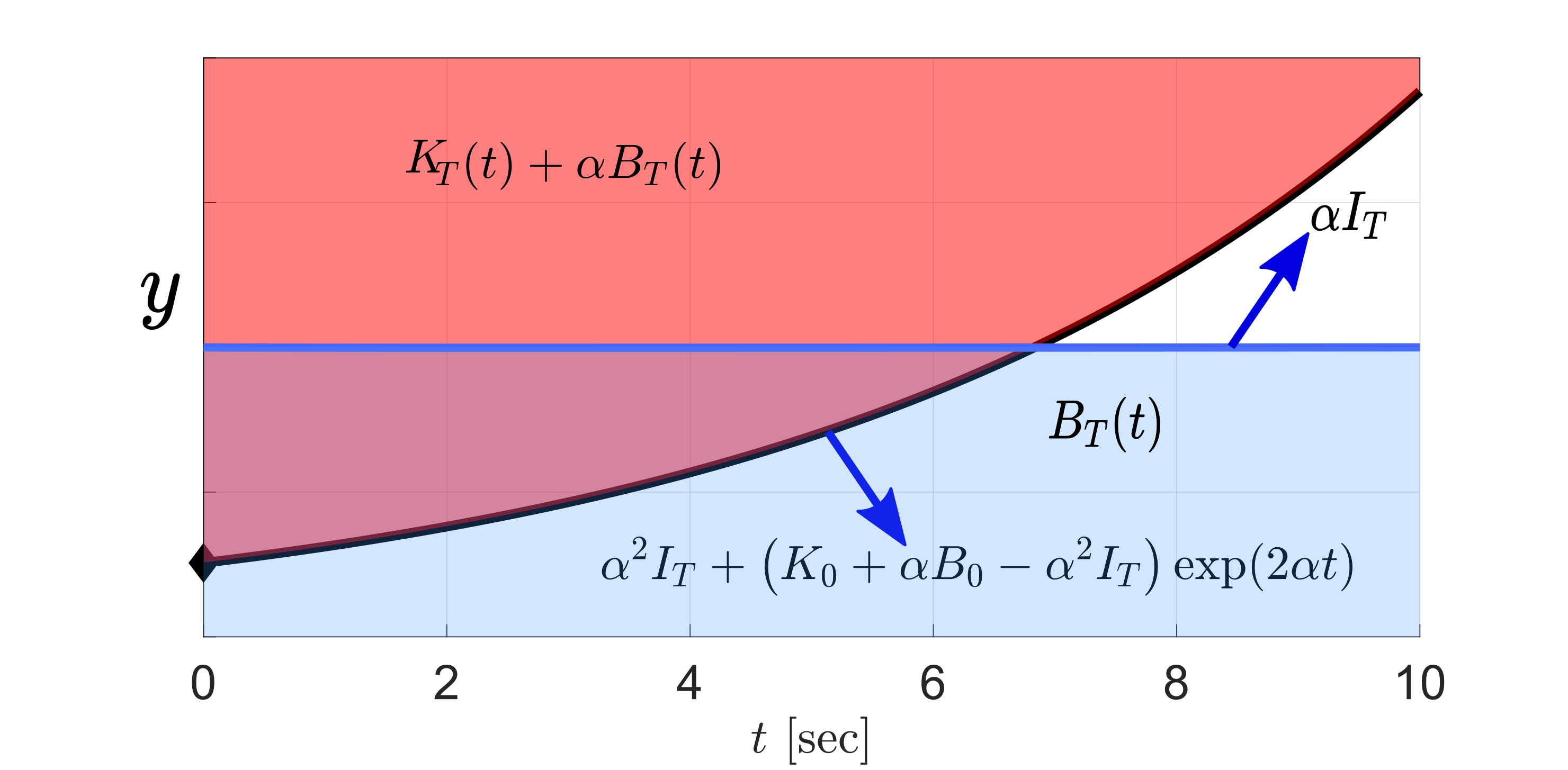} 
	\caption{\small{To ensure non-passive adversarial target dynamics for all $t>0$, $K_T(t) +  \alpha B_T(t)$ should belong to the red area above the curve $y=\alpha^2 I_T + \big( K_{0} + \alpha B_{0} - \alpha^2 I_T \big) \exp(2\alpha t)$ while $B_T(t)$ should belong to the blue area under the line $y=\alpha I_T$.}}
	\vspace{-2ex}
	\label{fig:svgSynth}
\end{figure}

Consequently, if the time-varying stiffness and damping profiles satisfy~\eqref{eq:ineq1} and~\eqref{eq:finalConst}, then the rate of change of the storage function $V_a(\cdot)$ satisfies~\eqref{eq:nonPassiveProof} and the adversarial target dynamics in~\eqref{eq:targetMain} become non-passive with respect to the input-output pair $(\dot{\theta}_{\text{d}},\, \alpha_{T_{\text{sw}}} \tau_{\text{sw}} + \alpha_{T_{\text{c}}} \tau_{\text{c}})$. As demonstrated in Figure~\ref{fig:svgSynth}, to ensure non-passive adversarial target dynamics for all $t>0$, $K_T(t) +  \alpha B_T(t)$ should belong to the red area above the curve $y=\alpha^2 I_T + \big( K_{0} + \alpha B_{0} - \alpha^2 I_T \big) \exp(2\alpha t)$ while $B_T(t)$ should belong to the blue area under $y=\alpha I_T$.

%
%

%
%
     \section{Closed-Loop Adaptive  Attack Policy}
\label{sec:attackDesign}

In this section we will design closed-loop adaptive attack policies that will make the angular position of the driver's input device to closely follow the adversarial angular position reference, which is generated by the non-passive  target dynamical system given by~\eqref{eq:targetMain}. Our adaptive control scheme follows the methodology proposed by Dawson and collaborators~\cite{baviskar2008adjustable}. 

We first define \emph{the driver experience error signal}, i.e., the angular position tracking error between the adversarial target dynamics generated reference $\theta_{\text{d}}$ and the driver input device angular position $\theta_{\text{sw}}$, as 
\begin{equation}
e_{\text{sw}} := \theta_{\text{d}} - \theta_{\text{sw}}.
\label{eq:errorSigsw}
\end{equation}
Furthermore, we define \emph{the locked tracking error signal}, i.e., the angular position tracking error between the driver input device and the steering column, as 
\begin{equation}
e_{\text{c}} := \theta_{\text{sw}} - \theta_{\text{c}}. 
\label{eq:errorSigc}
\end{equation}

Using the angular position tracking errors in~\eqref{eq:errorSigsw} and~\eqref{eq:errorSigc}, we also define the filtered error signals as 
\begin{equation}
r_{\text{sw}} := \dot{e}_{\text{sw}} + \mu_{\text{sw}} e_{\text{sw}},\, r_{\text{c}} := \dot{e}_{\text{c}} + \mu_{\text{c}} e_{\text{c}}, 
\label{eq:filteredErr}
\end{equation}
where the positive gains $\mu_{\text{sw}}$ and $\mu_{\text{c}}$ are design parameters. 

To state the closed-loop adaptive attack policy, we consider the regression matrices  
\begin{equation}
Y_{\text{sw}} := [Y_{N_{\text{sw}}}, -\tau_{\text{sw}}, \ddot{\theta}_{d1}  + \mu_{\text{sw}} \dot{e}_{\text{sw}}] \in \mathbb{R}^{1\times 4}, 
\label{eq:Ysw}
\end{equation}
and 
\begin{equation}
Y_{\text{c}} := [-Y_{N_{\text{sw}}}, \tau_{\text{sw}}, T_{\text{sw}}, Y_{N_{\text{c}}}, -\tau_{\text{c}}, \mu_{\text{c}}\dot{e}_{\text{c}}]\in \mathbb{R}^{1\times 8}, 
\label{eq:Yc}
\end{equation}
where  $Y_{N_{\text{sw}}}\in \mathbb{R}^{1\times 2}$ and $Y_{N_{\text{c}}}\in \mathbb{R}^{1\times 2}$ are the regression vectors given by the linearly parametrizable forms in~\eqref{eq:linParam}. Therefore, the row vectors $Y_{\text{sw}}$ and $Y_{\text{c}}$ consist of measurable quantities that the adversary is assumed to be able to read and/or accurately estimate from the in-vehicle communication network (IVN). As demonstrated in
numerous white-hat attack experiments, including a renowned  wireless attack against Tesla vehicles~\cite{nie2017free},
an adversary who has infiltrated the CAN bus of the target vehicle can read live data
 from the IVN (such as the target vehicle speed, its engine speed/torque, etc.).

In addition to the regression matrices given by~\eqref{eq:Ysw} and~\eqref{eq:Yc}, the \emph{unknown parameters} in the HSC steering dynamics in~\eqref{eq:dynMain} are utilized to define the following unknown vectors 
\begin{equation}
	\phi_{\text{sw}} := [\phi_{N_{\text{sw}}}, \alpha_{\text{sw}}, I_{\text{sw}}]^\top,
	\label{eq:phisw}
\end{equation}
\begin{equation}
	\phi_{\text{c}} := [\frac{I_{\text{c}}}{I_{\text{sw}}} \phi_{N_{\text{sw}}},  \frac{I_{\text{c}}}{I_{\text{sw}}} \alpha_{1}, \frac{I_{\text{c}}}{I_{\text{sw}}}, \phi_{N_{\text{c}}}, \alpha_{\text{c}}, I_{\text{c}}]^\top, 
	\label{eq:phic}
\end{equation}
where  $\phi_{N_{\text{sw}}}$ and $\phi_{N_{\text{c}}}$ are constant vectors in the linearly parametrizable forms given by~\eqref{eq:linParam}. 

\noindent{\textbf{Adversarial Adaptive Update Laws.}} It is assumed that the adversary does not have any knowledge about the constant vectors $\phi_{{\text{sw}}}$ and $\phi_{{\text{c}}}$ in~\eqref{eq:phisw} and~\eqref{eq:phic}, which capture the effect of vehicle physical parameters on the HSC steering dynamics. Rather, the adversary utilizes the estimated vectors ${\hat{\phi}}_{\text{sw}}$ and ${\hat{\phi}}_{\text{c}}$, whose evolution are governed according to the \emph{adaptive update laws}  
\begin{equation}
\dot{\hat{\phi}}_{\text{sw}} = \Gamma_{\text{sw}} Y_{\text{sw}}^\top r_{\text{sw}},\, \dot{\hat{\phi}}_{\text{c}} = \Gamma_{\text{c}} Y_{\text{c}}^\top r_{\text{c}}, 
\label{eq:adaptiveUpdate}
\end{equation}
where the diagonal matrices $\Gamma_{\text{sw}}$ and $\Gamma_{\text{c}}$ are positive definite and constant. As it can be seen from~\eqref{eq:adaptiveUpdate}, the update law dynamics depend on the filtered error signals $r_{\text{sw}}$ and $r_{\text{c}}$ in~\eqref{eq:filteredErr} as well as the regression matrices $Y_{\text{sw}}$ in~\eqref{eq:Ysw} and $Y_{\text{c}}$ in~\eqref{eq:Yc}.

\noindent{\textbf{Adversarial Closed-Loop  Adaptive Attack Policy.}} The adversary can utilize the estimates obtained from the adaptive update laws in~\eqref{eq:adaptiveUpdate} as well as the filtered error signals in~\eqref{eq:filteredErr} to define  attack policy inputs $T_{\text{sw}}$ and $T_{\text{c}}$ as  
\begin{equation}
T_{\text{sw}} = k_{\text{sw}} r_{\text{sw}} + Y_{\text{sw}} \hat{\phi}_{\text{sw}},\, T_{\text{c}} = k_{\text{c}} r_{\text{c}} + Y_{\text{c}} \hat{\phi}_{\text{c}}, 
\label{eq:Tattack}
\end{equation}
where $k_{\text{sw}}$ and $k_{\text{c}}$ are constant positive design parameters, and the vectors ${\hat{\phi}}_{\text{sw}}$ and ${\hat{\phi}}_{\text{c}}$, which are obtained from~\eqref{eq:adaptiveUpdate}, are the estimated values of the unknown constant vectors ${{\phi}}_{\text{sw}}$ and ${{\phi}}_{\text{c}}$ given by~\eqref{eq:phisw} and~\eqref{eq:phic}, respectively. 

As it can be seen from Figure~\ref{fig:steerBlock}, the attack policy inputs $T_{\text{sw}}$ and $T_{\text{c}}$ in~\eqref{eq:Tattack} are then applied by the adversary to the feedback servo motor on the driver's input device and the drive servo motor on the steering column of the target vehicle, respectively. The adaptive attack policy inputs $T_{\text{sw}}$ and $T_{\text{c}}$ enjoy several important properties including asymptotic convergence of tracking error signals $e_\text{sw}$ and $e_\text{c}$ to zero  under proper conditions (see, e.g., Theorem 1 in~\cite{baviskar2008adjustable}). It is remarked that the class of adaptive update laws in~\eqref{eq:adaptiveUpdate} only guarantee convergence of the angular position tracking errors to zero. However, convergence to the true values of these parameters is not a guaranteed feature of these adaptation rules, which rely on the linear parametrization property of the  underlying dynamics (see~\cite{baviskar2008adjustable}  and~\cite[Chapter 9]{spong2020robot} for further details). 

\noindent{\textbf{Summary of the Adversarial Attack Policy Against the HSC Steering System.}} Having infiltrated the HSC steering control system, the adversary utilizes the linear combination $\alpha_{T_{\text{sw}}} \tau_{\text{sw}} + \alpha_{T_{\text{c}}} \tau_{\text{c}}$ consisting of the driver's input and the steering column input torques as the input to the properly designed non-passive and time-varying impedance target dynamics given by~\eqref{eq:targetMain} (see  Figure~\ref{fig:steerTargetDyn}).  Having computed the reference command $\theta_d$ from these target dynamics, the adversary computes the adaptive attack policy inputs $T_{\text{sw}}$ and $T_{\text{c}}$ from~\eqref{eq:Tattack} and applies them to the servo motors on the driver's and the steering column sides, respectively (see Figure~\ref{fig:steerBlock}). These closed-loop attack policies ensure that the driver input device angular position $\theta_{\text{sw}}$ closely follows   the malicious reference command $\theta_\text{d}$ generated by the non-passive target dynamics. 
     \section{Simulation Results}
\label{sec:sims}
In this section we present numerical simulation results to demonstrate the effectiveness of the proposed attack. The simulation parameters are given in Table~\ref{tab:TableSim}, where the HSC steering system  parameters are directly adopted from~\cite{baviskar2008adjustable}. 

	\begin{table}[t]
	\centering
	\begin{tabular}{|c|c|c|c|}
		\hline
		\rowcolor[HTML]{93C47D} 
		{\color[HTML]{93C47D} } 	Variable			 &   Value &  	Variable  & Value   \\ \hline
		$I_\text{sw}$ &  $1.16\times 10^{-2}$ kg.m\textsuperscript{2} &   $I_\text{c}$  & $2.35\times 10^{-2}$ kg.m\textsuperscript{2}  \\ 
		$B_\text{sw}$ &  $1.9\times 10^{-2}$ kg.m\textsuperscript{2}/s& 	$B_\text{c}$ &  $6 \times 10^{-2}$  kg.m\textsuperscript{2}/s\\ 
		$K_\text{sw}$ & $0$ N.m &      $K_\text{c}$ &  $0$ N.m  \\ 
		$\alpha_\text{sw}$ & $1$ &      $\alpha_\text{c}$ &  $1$  \\ 
		$\alpha_{T_{\text{sw}}}$ & $1$ &      $\alpha_{T_{\text{c}}}$ &  $0.15$  \\ 
		$\gamma$ & $0.02$ &      $C_d$ &  $150$ N.m  \\ \hline
	\end{tabular}
	\caption{\label{tab:TableSim} \small Numerical simulation parameters and their values.}
\end{table} 

Due to the tire–road interface reaction forces, a reaction torque will be exerted on the directional control assembly (see the block diagram in Figure~\ref{fig:steerBlock}). In the simulations, we have assumed the following functional form for the reaction torque $\tau_{\text{c}}$ (see, e.g.,~\cite{setlur2006trajectory}) 
\begin{equation}
	\tau_{\text{c}} = -C_d \tanh(\gamma \theta_{\text{c}}), 
	\label{eq:tauCForm}
\end{equation}
\noindent where the values of $\gamma$ and $C_d$ are given in Table~\ref{tab:TableSim}. 

\begin{figure}[t]
	\centering
	\vspace{0.1ex}
	\includegraphics[height=0.25\textwidth]{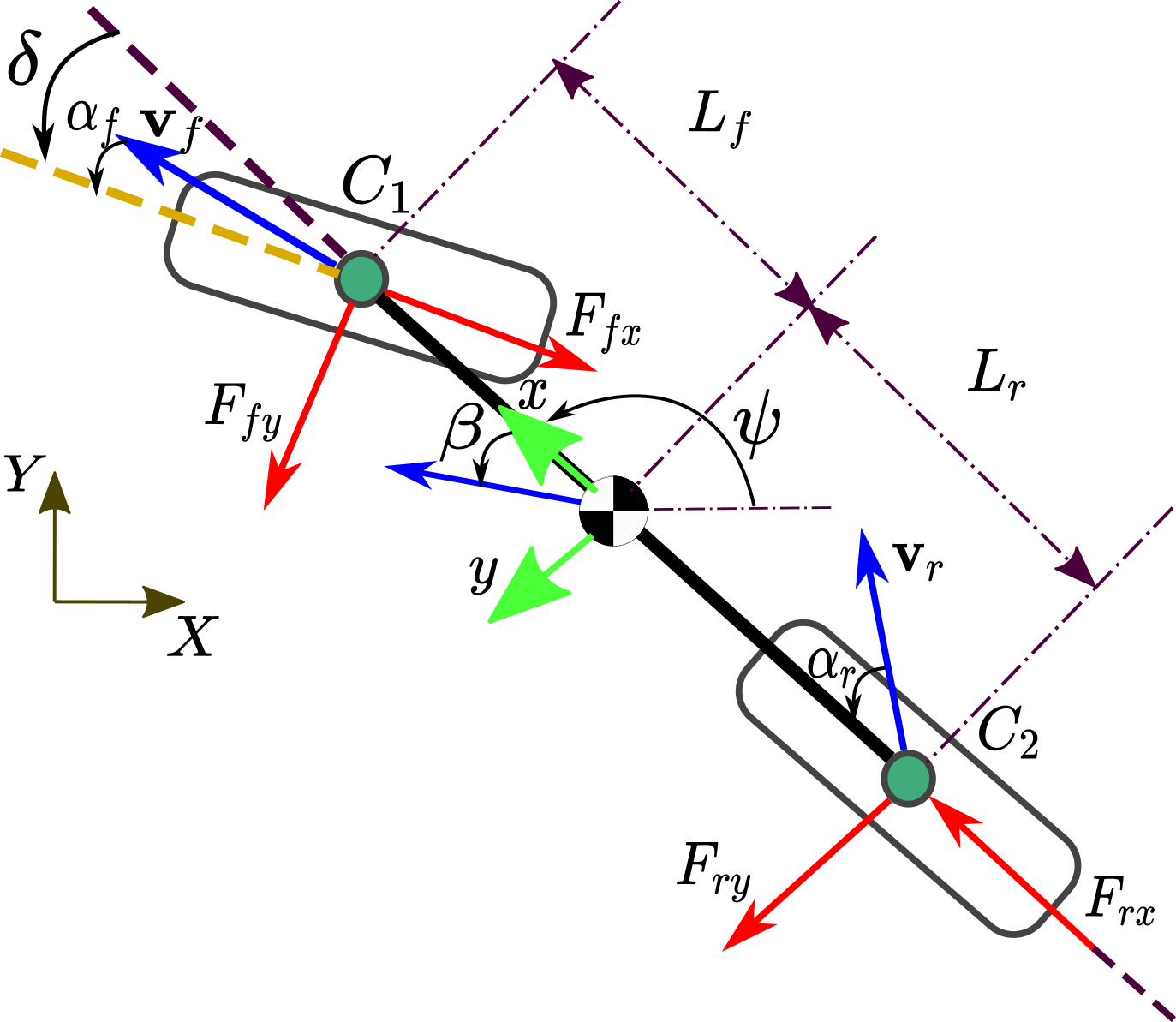}
	\vspace{-0.1ex}
	\caption{\small Bicycle model schematic.}
	\label{fig:bikeModel} 
\end{figure}

In addition to the HSC steering system dynamics,   we utilize a bicycle model for the vehicle dynamics (see Figure~\ref{fig:bikeModel}) to study the impact of the proposed attacks on the trajectory of a vehicle that is performing an obstacle clearing maneuver. In the bicycle model, we have  employed a hybrid physical/dynamic tire/road friction model.  The readers are referred to~\cite{li2013hybrid} for further details about this modeling approach. 

\begin{figure}[t]
	\includegraphics[height=0.26\textwidth]{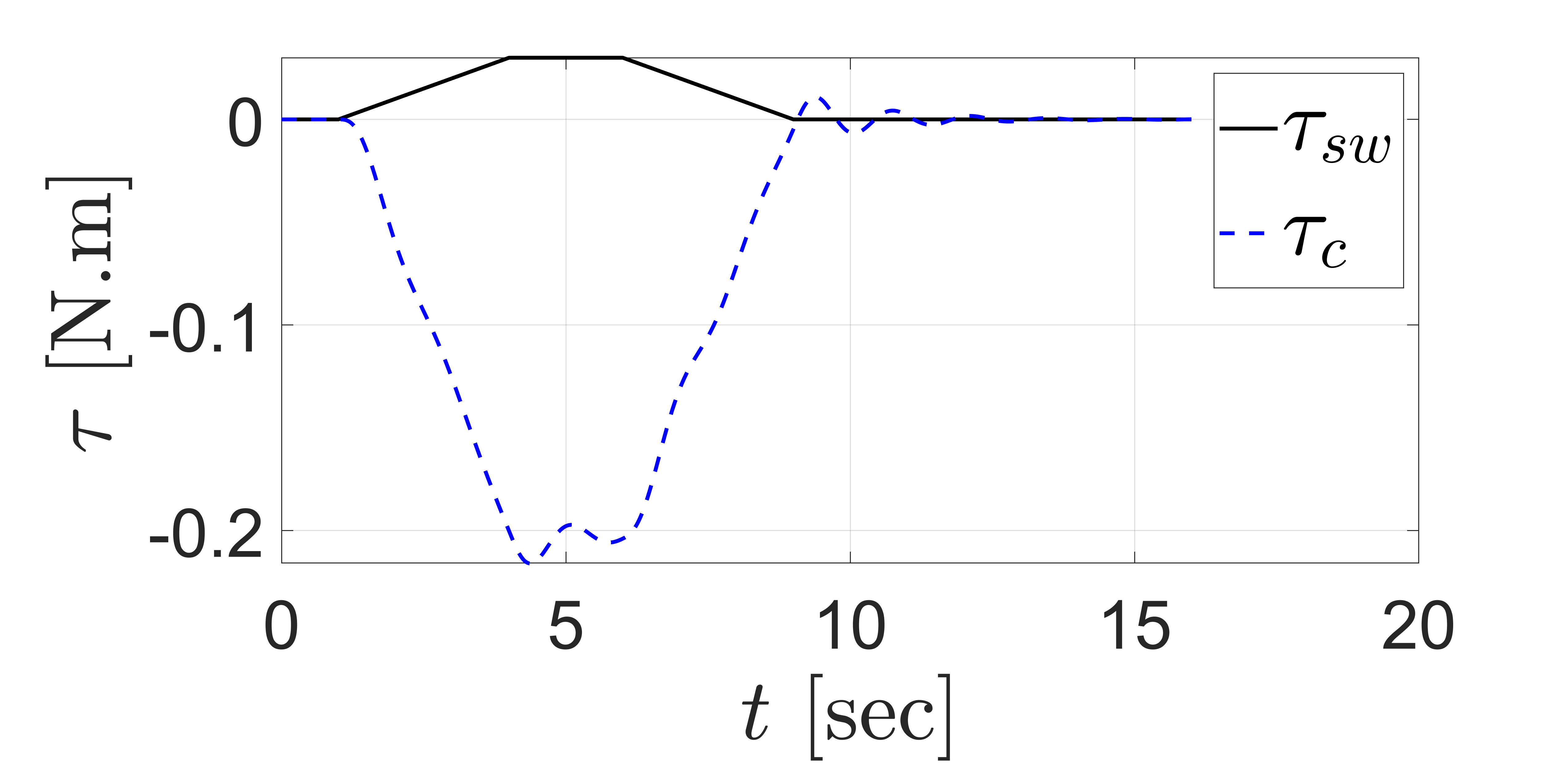} 
	\caption{\small The torque time profiles of the nominal (no attack) obstacle clearing maneuver using the HSC steering system: the human driver's (black)  and the steering column (dashed blue) torque time profiles.}
	\vspace{-1ex}
	\label{fig:Nominal1}
\end{figure}

\begin{figure}[t]
	\centering
	\includegraphics[width=0.5\textwidth]{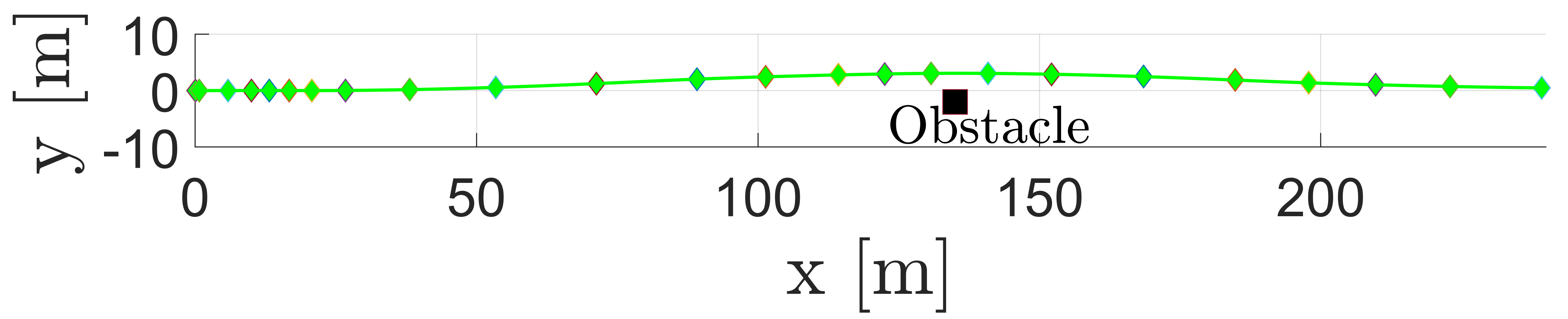} 
	\caption{\small The trajectory of the vehicle under the nominal (no attack) obstacle clearing maneuver.}
	\vspace{-1ex}
	\label{fig:Nominal3}
\end{figure}

\begin{figure}[t]
	\includegraphics[height=0.25\textwidth]{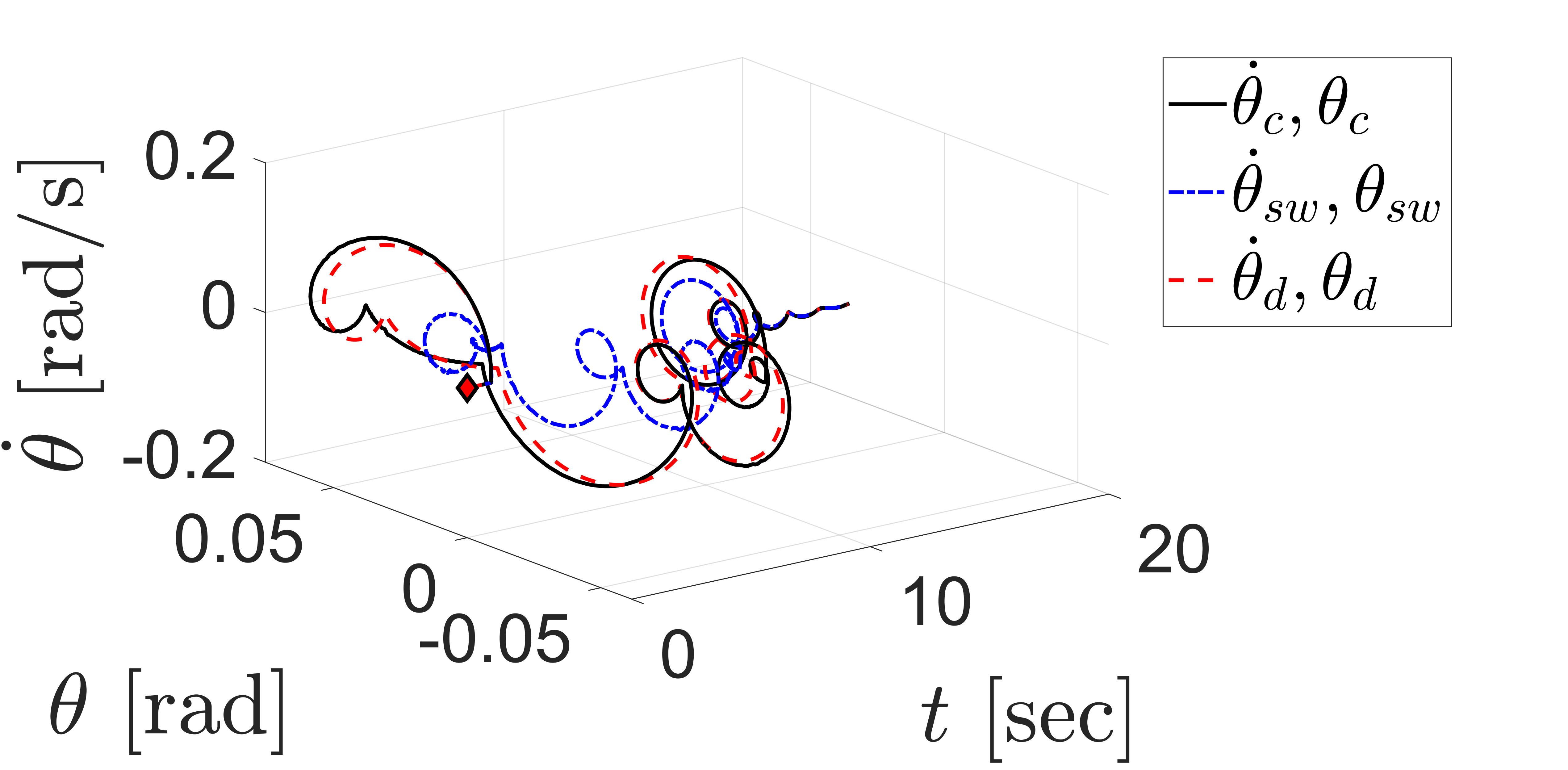} 
	\caption{\small The nominal (no attack) HSC steering system angular position/speed time profiles while performing an obstacle clearing maneuver: (black) the steering column angular time profiles; (dashed blue) the driver's input device angular time profiles; (dashed red) the angular time profiles generated by the nominal target dynamics in~\cite{baviskar2008adjustable}.}
	\vspace{-1ex}
	\label{fig:Nominal2}
\end{figure}

In the first collection of numerical simulations, we assume that there exists no cyber-physical attack on the HSC steering control system while the driver is performing a simple obstacle clearing maneuver. The nominal HSC steering control system is utilizing the asymptotically stable control schemes originally proposed by~\cite{baviskar2008adjustable}. Figures~\ref{fig:Nominal1} and~\ref{fig:Nominal2} depict the nominal HSC steering system torque  and angular position/speed time profiles, respectively. Figure~\ref{fig:Nominal3} depicts the trajectory of the vehicle under the nominal obstacle clearing maneuver.
 

In the second collection of numerical simulations under the proposed time-varying impedance attack against the target vehicle HSC steering system, we have assumed that the adversarial inertia and time-varying stiffness and damping profiles are given by $I_T=1\times 10^{-2}$ kg.m\textsuperscript{2}, $K_T(t)=2.8\times 10^{-2}\exp(2.1\alpha t)$ N.m, where $\alpha=0.5$, and $B_T(t)=4.99\times 10^{-3}$ kg.m\textsuperscript{2}/s, respectively. These time-varying stiffness and damping time profiles satisfy~\eqref{eq:ineq1} and~\eqref{eq:finalConst}, which guarantee a non-passive adversarial target dynamics. Furthermore, we have considered the weighting factors  $\alpha_{T_{\text{sw}}}$ and $\alpha_{T_{\text{c}}}$ in~\eqref{eq:targetMain} to be equal to $0.15$ and $1$, respectively. The parameters of the closed-loop adaptive attack policy inputs $T_{\text{sw}}$ and $T_{\text{c}}$ and their update laws in~\eqref{eq:adaptiveUpdate} and~\eqref{eq:Tattack} are chosen to be  $\mu_{\text{sw}}=\mu_{\text{c}}=1.01$, $k_{\text{sw}}=k_{\text{c}}=1$, $\Gamma_{\text{c}}=80\mathbf{I}_8$, and $\Gamma_{\text{sw}}=80\mathbf{I}_4$, respectively. Note that we denote the identity matrix of size $N$ by $\mathbf{I}_N$. 

\begin{figure}[t]
	\includegraphics[height=0.25\textwidth]{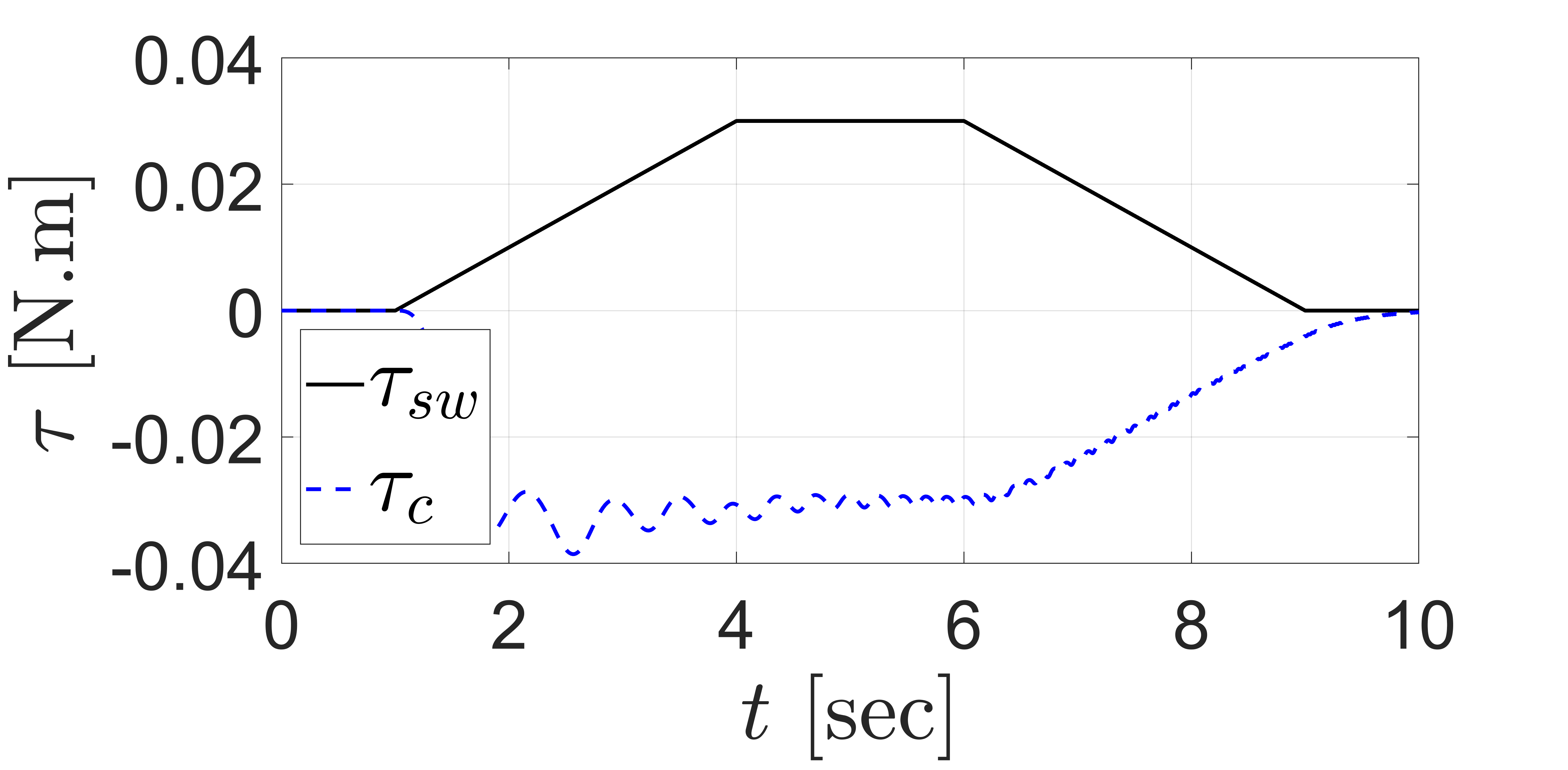} 
	\caption{\small The torque time profiles of the obstacle clearing maneuver using the HSC steering system under attack: the human driver's (black)  and the steering column (dashed blue) torque time profiles.}
	\vspace{-1ex}
	\label{fig:unstable1}
\end{figure}

\begin{figure}[t]
	\includegraphics[height=0.25\textwidth]{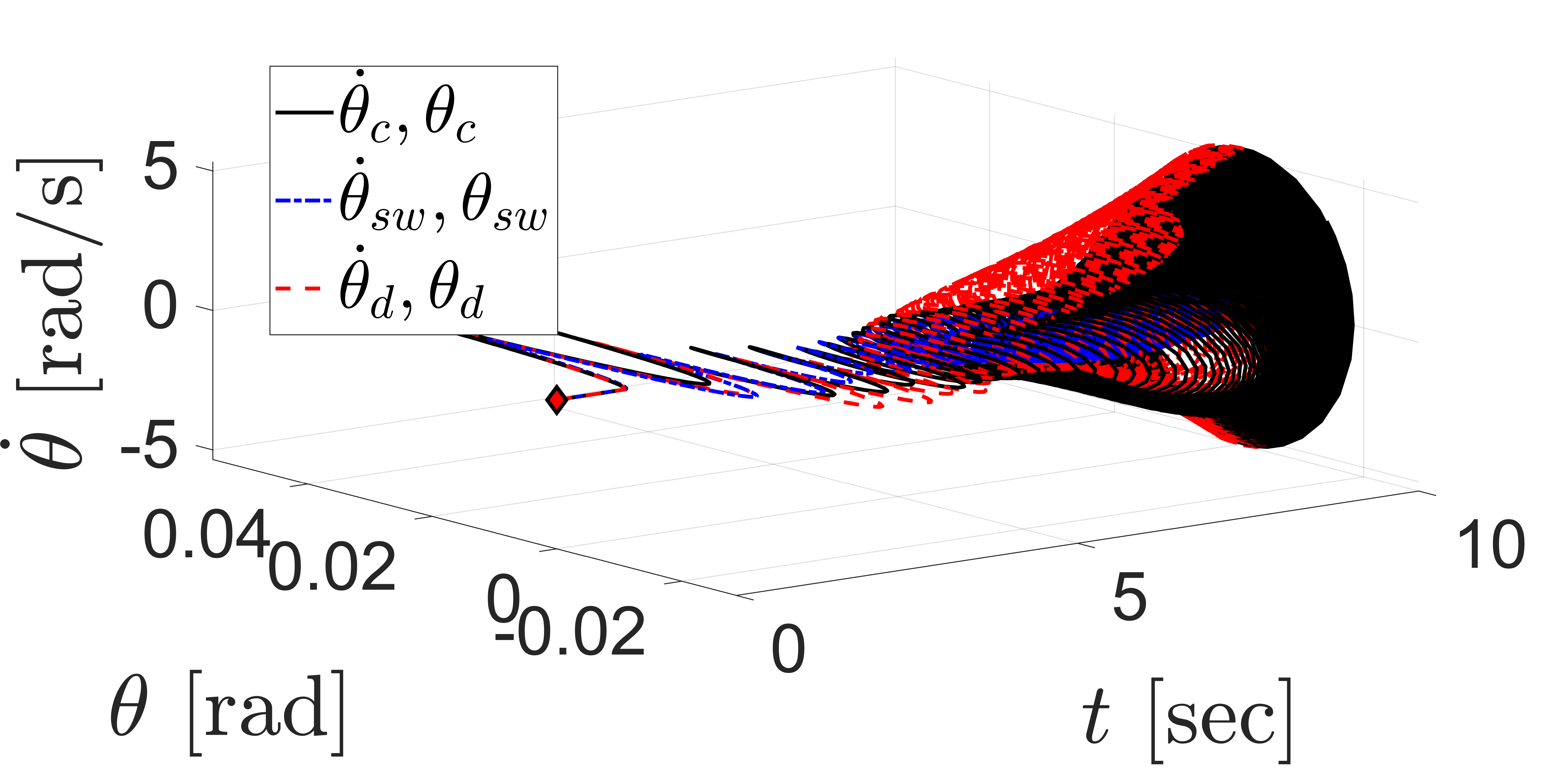} 
	\caption{\small The HSC steering system angular position/speed time profiles under the proposed time-varying impedance attack: (black) the steering column angular time profiles; (dashed blue) the driver's input device angular time profiles; (dashed red) the angular time profiles generated by the adversarial target dynamics given by~\eqref{eq:targetMain}.}
	\vspace{-1ex}
	\label{fig:unstable2}
\end{figure}

Figures~\ref{fig:unstable1} and~\ref{fig:unstable2} depict the attacked HSC steering system torque  and angular position/speed time profiles, respectively. As it can be seen from Figure~\ref{fig:unstable2}, the target dynamics  are injecting energy into the driver input device and the steering column systems by providing the closed-loop adaptive control policy with angular reference commands that get modulated in response to the driver's input torque and the steering column torque. Additionally, from the same figure, it can be seen that the closed-loop adaptive control policy is making the driver's input device and the steering column angular positions to closely follow the malicious reference command $\theta_d$ generated by the adversarial target dynamics in~\eqref{eq:targetMain}. 

\begin{figure}[t]
	\centering
	\includegraphics[width=0.5\textwidth]{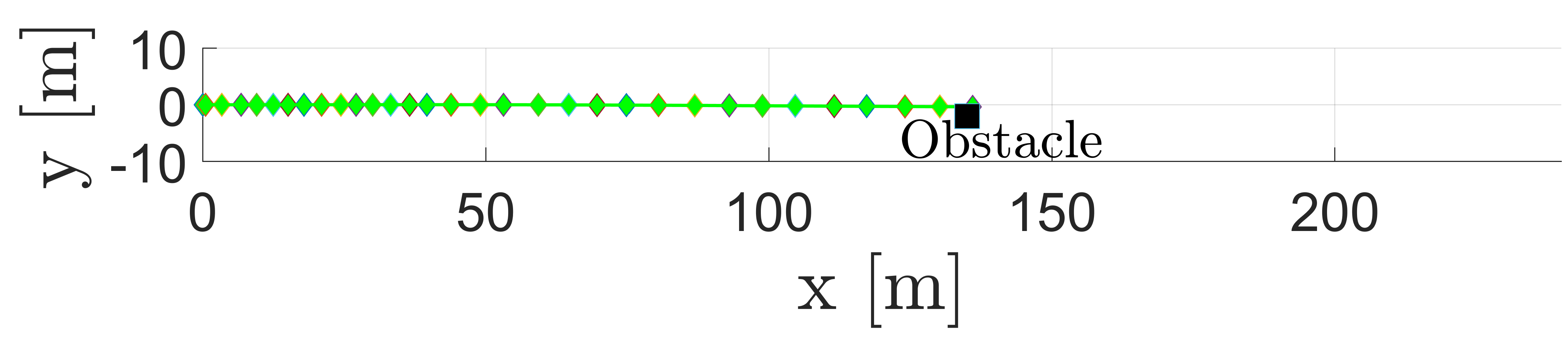} 
	\caption{\small The trajectory of the vehicle under the proposed time-varying impedance attack resulting in collision with the obstacle.}
	\vspace{-1ex}
	\label{fig:unstable3}
\end{figure}

Figure~\ref{fig:unstable3} depicts the trajectory of the vehicle when the obstacle clearing maneuver is taking place under the proposed time-varying impedance attack input. As it can be seen from this figure, the driver fails to clear the obstacle and the targeted vehicle collides with the obstacle. Intuitively, this collision can be explained by noticing that the attacked HSC steering system dynamics are stiffening  under the influence of the time-varying stiffness $K_T(t)=2.8\times 10^{-2}\exp(2.1\alpha t)$ in the adversarial target dynamics. This is why the attacked HSC steering system cannot make a proper response to the human driver's input in contrast with the nominal obstacle clearing maneuver depicted in Figure~\ref{fig:Nominal3}. This stiffening effect can also be observed when comparing the nominal steering column torque time profile $\tau_{c}(t)$ in Figure~\ref{fig:Nominal1} with its counterpart (under attack scenario) in Figure~\ref{fig:unstable1}. 

\begin{figure}[t]
	\includegraphics[width=0.5\textwidth]{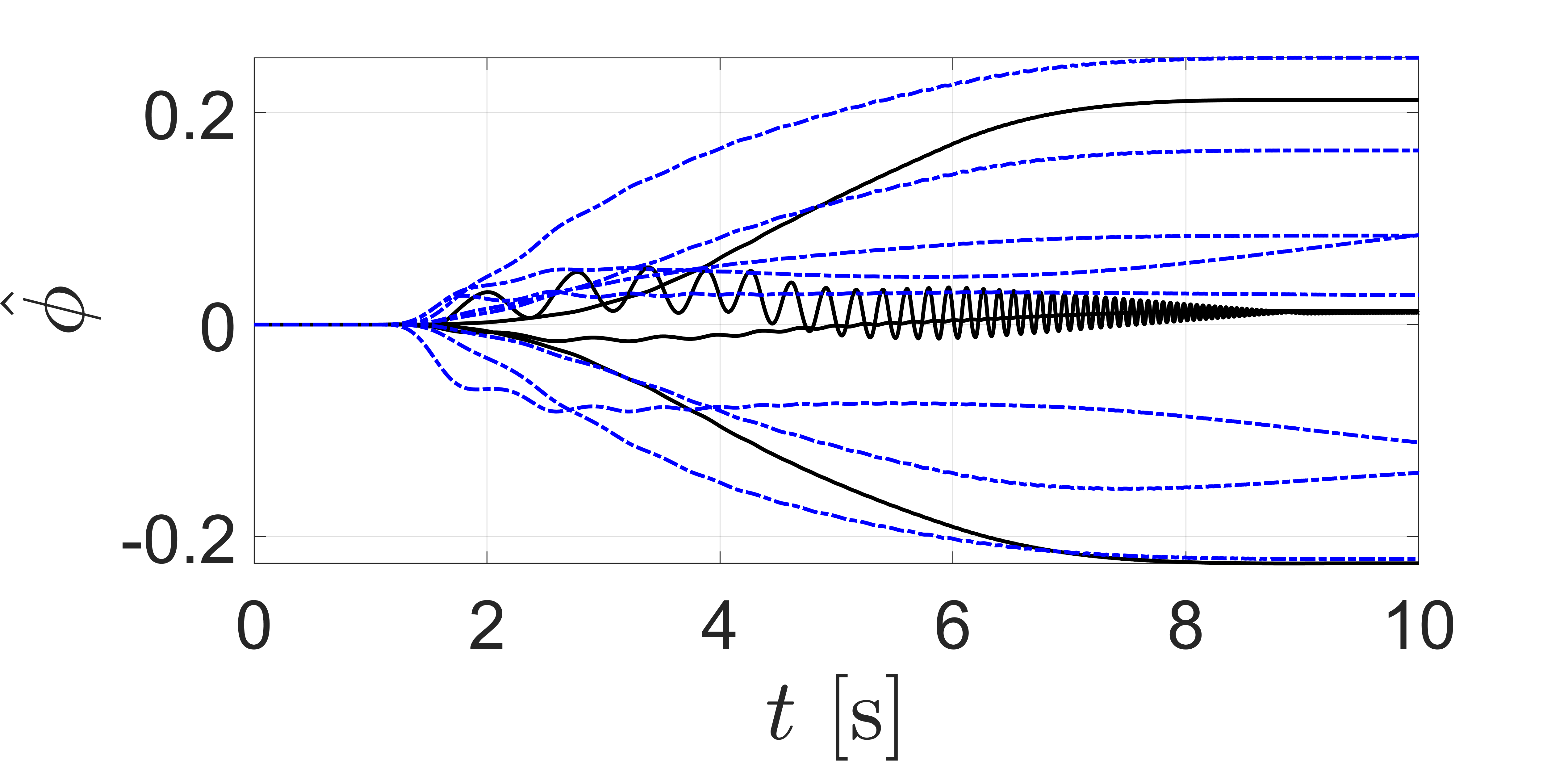} 
	\caption{\small The time evolution of the elements of the estimated vectors  $\hat\phi_{\text{sw}}$ and  
		$\hat\phi_{\text{c}}$ under the adversarial adaptive update laws in~\eqref{eq:adaptiveUpdate}.}
	\vspace{-1ex}
	\label{fig:unstable4}
\end{figure}

Figure~\ref{fig:unstable4} depicts the evolution of the elements of the estimated vectors  $\hat\phi_{\text{sw}}$ and  $\hat\phi_{\text{c}}$ under the adversarial adaptive update laws given by~\eqref{eq:adaptiveUpdate}. It should be noted the class of adaptive update laws, which rely on the linear parametrization property of the HSC steering dynamics, only guarantee convergence properties of angular position tracking errors. However, convergence to the true values of the unknown parameter vectors $\phi_{\text{sw}}$ and  $\phi_{\text{c}}$ is not a guaranteed feature of these adaptation rules.
     \section{Concluding Remarks and Future Research Directions}
\label{sec:conc}
The aim of this study was to explore the potential of adversarial techniques in disrupting the interaction dynamics between human drivers and vehicle haptic shared control (HSC) steering systems. While previous research in robotics has focused on ensuring stability and passivity in human-robot interaction dynamics, this paper took a different approach. It proposed the synthesis of time-varying impedance profiles that would create a non-passive interaction between the human driver and the HSC steering system. The study highlighted the importance of HSC steering systems in critical safety scenarios, such as facilitating smooth transitions from automation to human control and reducing conflicts between drivers and vehicles. Through practical experiments, this paper demonstrated the physical capabilities of adversaries who aim to destabilize the stability of driver-vehicle interaction. Future research will include modeling the neuromuscular dynamics of the drivers in reaction to such attacks and devising real-time passivity monitoring algorithms (see, e.g.,~\cite{welikala2022line}) for detection of such attacks against the vehicle HSC steering system.  
\section*{ACKNOWLEDGMENT}
This work is supported by NSF Award CNS--2035770.     
\bibliographystyle{./styles/IEEEtran}
\bibliography{SaTCBib}
\end{document}